\documentstyle[prl,aps,twocolumn,psfig]{revtex}
\begin{document}
\title{Quantum entanglement in the NMR
implementation of the Deutsch-Jozsa 
algorithm}
\author{Arvind$^1$~\cite{arv-email},
Kavita Dorai$^2$~\cite{kav-email},  
and Anil Kumar$^{2,3}$~\cite{ak-email}}
\address{$^1$Department of Physics, 
Guru Nanak Dev University, Amritsar 143005 India}
\address{$^2$Department of Physics, and 
$^3$Sophisticated Instruments Facility,
Indian Institute of Science, Bangalore 560012 India}
\maketitle 
\begin{abstract}
A scheme to execute an $n\/$-bit Deutsch-Jozsa
(D-J) algorithm using $n\/$ qubits has been implemented 
for up to three qubits on an NMR quantum computer.  
For the one and two bit Deutsch problem, the qubits do 
not get entangled, hence the NMR  implementation is
 achieved without using spin-spin interactions.
It is for the three bit case, that the manipulation 
of entangled states becomes essential.  The interactions 
through scalar J-couplings in NMR spin systems have been
exploited to implement entangling transformations
required for the three bit D-J algorithm.
\end{abstract}
The utilization of the intrinsically quantum
mechanical nature of the physical world to
widen the scope of computational algorithms is
one of the important discoveries of this 
decade~\cite{divin-sc-95,lloyd-sc-95,chua-sc-95}.
It was shown recently that quantum computers can
perform certain computational tasks exponentially
faster than classical 
computers~\cite{deu-roy-92,shor-siam-97,grover-prl-97}. 

The Cleve version of the D-J algorithm~\cite{cleve-royal-98}, 
which requires $n+1\/$ qubits to solve the
$n$-bit Deutsch problem, has been implemented
by several research groups using 
NMR~\cite{ch-nature-98,jones-jcp-98,lin-cpl-98,kavita-qph}.
It has been shown recently that the  $n\/$-bit Deutsch
problem can be solved using $n\/$ 
qubits alone~\cite{collins-pra-98}.

In this paper, we experimentally demonstrate that 
the $n\/$-bit D-J algorithm does not require $n+1\/$ 
qubits for its implementation. By doing away with
the extra qubit, the algorithm can be more easily 
accessed for a greater number of qubits. 
Furthermore, the one-bit and 
two-bit implementations of the
modified version do not involve 
quantum entanglement~\cite{collins-pra-98,ekert-amj-94}. 
In these cases, only the concept of coherent 
superposition is exploited, to prepare
``in parallel'' an input state which is a superposition of all
possible classical inputs, and the experiment has been
performed without using spin-spin interactions.
It is only in the implementation  
for three or more qubits does quantum entanglement 
play a vital role. A judicious combination of rf pulses
and free evolution intervals (under the interaction
Hamiltonian), has been employed to construct
the required entangling transformations.

Consider an $n\/$-bit binary string $x\/$; a  
function $f\/$ can be defined on this $n\/$-bit
domain space to a 1-bit range space, with the
restriction that either the output is the same
for all inputs (the function is constant) or the
output is $0\/$ for half the inputs and $1\/$ for
the other half (the function is balanced).  All
the $2^n\/$ possible input strings are valid
inputs for the function ($f(x)=\{0,1\}\/$). 
In quantum computation, these
$n\/$-bit logical strings are in one-to-one
correspondence with the eigen states of
$n\/$-qubits, and one can hence label the logical
string $x\/$ by the eigenstate $\vert x
\rangle\/$.  Classically, for an
$n\/$-bit domain space, one needs to compute the
function at least $2^{n/2}+1\/$ times  in order to
determine whether it is constant or balanced.
The D-J algorithm achieves this on a quantum computer using
only a single function
call~\cite{deu-roy-92,cleve-royal-98}.

The usual implementation of the D-J algorithm
for $n\/$ bits requires $n+1\/$ qubits,  
the function $f\/$ being encoded
through an $f\/$-dependent unitary
transformation, 
\begin{equation}
\vert x \rangle_{\mbox{\tiny n-bit}} 
\vert y \rangle_{\mbox{\tiny 1-bit}}
\stackrel{U_f}{\longrightarrow}
\vert x \rangle_{\mbox{\tiny n-bit}}
\vert y \oplus f(x) \rangle_{\mbox{\tiny 1-bit}}
\label{function-call-old}
\end{equation}
where $\oplus\/$ denotes addition modulo $2\/$.
The implementation of the unitary transformation
$U_f\/$, along
with the Hadamard transformation, then suffices
to distinguish the function as constant or 
balanced~\cite{deu-roy-92,cleve-royal-98}.
A Hadamard transformation on one qubit 
mixes the eigenstates maximally,  
\begin{equation}
\begin{array}{c}
\vert 0\rangle \stackrel{H}{\rightarrow} 
\frac{\scriptstyle 1}{\scriptstyle \sqrt{2}}
(\vert 0 \rangle + \vert 1 \rangle) \\
\vert 1\rangle \stackrel{H}{\rightarrow} 
\frac{\scriptstyle 1}{\scriptstyle \sqrt{2}}
(\vert 0 \rangle - \vert 1 \rangle)
\end{array}
;\,
H = H^{-1}=\frac{1}{\sqrt{2}}
\left(\begin{array}{lr} 
{ 1} & {1}\\
{ 1} & {-1}
\end{array}
\right) 
\label{hadamard}
\end{equation}
The Hadamard transformation for $n$-qubits 
is the tensor product of the one-qubit
transformation ($H^n = H \otimes H
\otimes H \cdots \otimes H\/$), its action on
the $n$-qubit eigen states being 
\begin{equation}
H^n \vert x \rangle = \sum_{y=0}^{2^n -1}
(-1)^{\oplus \sum_{j} x_j y_j} \vert y \rangle
\end{equation}
where $x_j\/$ and $y_j\/$ are the $j$th
entries of the $n$-bit strings $x$ and $y$.

A modified scheme can be designed to solve
the $n$-bit Deutsch problem, using 
$n$ qubits alone~\cite{collins-pra-98}. 
Here, for every function $f\/$ a unitary
transformation is constructed,  such that its
action on the eigenstates of $n$-qubits is 
\begin{equation}
\vert x \rangle_{\mbox{\tiny n-bit}}
\stackrel{U_{f}}{\longrightarrow}
(-1)^{f(x)} \vert x \rangle_{\mbox{\tiny n-bit}}
\label{call-mech}
\end{equation}

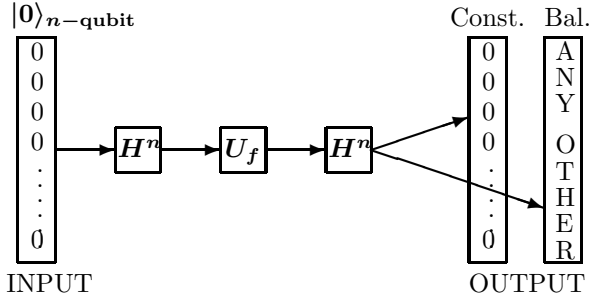
\begin{figure}
\unitlength=1mm
\begin{picture}(75,40)
\thicklines
\multiput(5,20)(14,0){3}{\vector(1,0){8}}
\put(47,20){\vector(3,1){13}}
\put(47,20){\vector(3,-1){23}}
\multiput(13,17)(14,0){3}{\line(1,0){6}}
\multiput(13,23)(14,0){3}{\line(1,0){6}}
\multiput(13,17)(14,0){3}{\line(0,1){6}}
\multiput(19,17)(14,0){3}{\line(0,1){6}}
\multiput(0,5)(60,0){2}{\line(1,0){5}}
\multiput(0,35)(60,0){2}{\line(1,0){5}}
\multiput(0,5)(60,0){2}{\line(0,1){30}}
\multiput(5,5)(60,0){2}{\line(0,1){30}}
\multiput(70,5)(0,30){2}{\line(1,0){5}}
\multiput(70,5)(5,0){2}{\line(0,1){30}}
\put(1.75,7){0}
\multiput(2.5,7)(0,2){6}{.}
\multiput(1.75,20)(0,4){4}{0}
\put(61.75,7){0}
\multiput(62.5,7)(0,2){6}{.}
\multiput(61.75,20)(0,4){4}{0}
\boldmath
\multiput(13.3,19)(28,0){2}{$H^n$}
\put(27.5,19){$U_f$}
\put(-1,37){$\vert 0 \rangle_{n-{\rm qubit}}$}
\unboldmath
\put(71.5,32){A}
\put(71.5,28.5){N}
\put(71.5,25){Y}
\put(71.5,19.5){O}
\put(71.5,16){T}
\put(71.5,12.5){H}
\put(71.5,9){E}
\put(71.5,5.5){R}
\put(-1.5,1){INPUT}
\put(60,1){OUTPUT}
\put(57.5,36.5){Const.}
\put(70,36.5){Bal.}
\end{picture}
\vspace{12pt}
\caption{The block diagram for the modified D-J algorithm.}
\label{working}
\end{figure}

Consider $n$ qubits, all in the state 
$\vert 0 \rangle\,$; a Hadamard transformation
$H^n\/$ converts this state to a linear
superposition of all $2^n\/$ eigenstates
with equal amplitudes and no phase differences. 
The unitary transformation $U_f\/$ (defined in
Eqn.~\ref{call-mech}) acting on this
state, introduces an $f$-dependent phase
factor in each eigenstate in the superposition.
 At this juncture, all
information about $f\/$ is encoded in the
quantum state of the $n$ qubits. A
Hadamard transformation $H^n\/$ is once
again applied in order to extract the function's
constant or balanced nature:  
\begin{eqnarray}
\vert 0 \rangle \stackrel{H^n}{\longrightarrow}
\sum_{x=0}^{2^n-1}\vert x \rangle
\stackrel{U_f}{\longrightarrow}
\sum_{x=0}^{2^n-1}
(-1)^{f(x)} \vert x \rangle 
\stackrel{H^n}{\longrightarrow} \quad\quad&&
\nonumber \\
\sum_{x=0}^{2^n-1}
\sum_{y=0}^{2^n-1}
(-1)^{f(x)} (-1)^{\oplus \sum_j x_j y_j} 
\vert y \rangle &&
\label{working-eq}
\end{eqnarray}
The final expression 
for the output state in Eqn.~\ref{working-eq}
has an amplitude $1\/$ for the state
$\vert 0 \rangle_{\mbox{\tiny n-bit}}\/$ for
a constant function and an amplitude $0\/$
for a balanced function.
The categorisation of the function  as constant
or balanced  through a single function call using
$n\/$ qubits, 
is shown pictorially
in Fig.~\ref{working}.
The number of functions for the $n$-bit Deutsch
problem is ${}^{N}C{}_{N/2} + 2\/$ (where
$N = 2^{n}\/$).   The experimental 
implementation of the modified D-J algorithm for 
$n\/$ bits requires the realisation of the  
unitary transformation corresponding to each
of these functions, and the $n$-bit Hadamard
transformation,  on a physical system. 
We now proceed towards the NMR
implementation of the modified D-J algorithm for
one, two and three qubits, where the number of
functions are 4, 8, and 72 respectively. 

The pseudo-Hadamard transformation~\cite{jones-jmr-98}  
(practically equivalent to the Hadamard operator 
described in Eqn.~\ref{hadamard}) achieved by 
a $(90^{0})_{y}\/$ pulse on a spin, has
been utilised in our experiments.  For the case of
two and three qubits, the same has been
achieved by a $(90^{0})_{y}\/$ pulse applied
non-selectively on all the spins. 

The $n$-bit unitary transformations $U_f\/$ corresponding
to the functions $f\/$, are diagonal in the eigenbasis 
and find a natural description in terms of the 
single-spin operators, 
\begin{equation}
I^{(j)}
= \left( \begin{array}{cc}
1 & 0 \\
0 & 1
\end{array}
\right)
\,\,\,\,
;
\,\,\,\,
\sigma^{(j)}_{z} = \left(
 \begin{array}{cc}
1 & 0 \\
0 & -1
\end{array}
\right)
\end{equation}
where $j\/$ labels the qubit involved. 
The action of $U_f\/$ on an eigenstate (as described
in Eqn.~\ref{call-mech}), can been used to calculate
the explicit matrix forms of $U_f\/$, for every
function $f\/$.
 
\begin{figure}[h]
\unitlength=1mm 
\vspace*{-0.5cm}
\begin{picture}(80,55)
\put(0,3){\psfig{figure=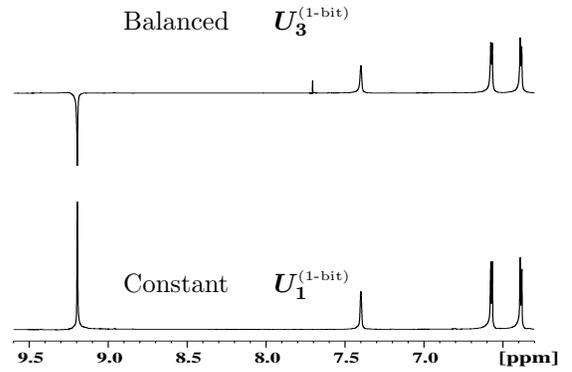,height=4.5cm,width=7cm,angle=0}}
\boldmath
\put(15,13){Constant} 
\put(35,13){$U^{\mbox{\tiny (1-bit)}}_{1}\/$}
\put(15,48){Balanced}
 \put(35,48){$U^{\mbox{\tiny (1-bit)}}_{3}\/$}
\unboldmath
\end{picture}
\caption{The modified D-J algorithm for one qubit implemented
on 5-nitro-2-furaldehyde, the proton resonating
at 9.2 ppm being chosen as the single qubit.} 
\label{onebit}
\end{figure}

The  operator representations of the four unitary 
transformations for the one-bit modified D-J algorithm 
are
\begin{equation}
\begin{array}{ccll}
U^{\mbox{\tiny (1-bit)}}_{1} &=& I^{(1)}&
\quad \mbox{(Const.)}  \\
U^{\mbox{\tiny (1-bit)}}_{2} &=& - I^{(1)} &
\quad \mbox{(Const.)}  \\
U^{\mbox{\tiny (1-bit)}}_{3} &=&  \sigma^{(1)}_{z} &
\quad \mbox{(Bal.)}  \\
U^{\mbox{\tiny (1-bit)}}_{4} &=& -\sigma^{(1)}_{z}& 
\quad \mbox{(Bal.)}  
\end{array}
\end{equation}
A pseudo-Hadamard operation achieved by a  
$(90^{0})_{y}\/$ pulse is applied on a thermal
initial state, in order to create a coherent 
superposition prior to
applying the desired unitary transformations $U_f\/$.
The constant functions correspond to a 
``do-nothing'' operation, while the balanced
functions are achieved by a rotation by the
angle $\pi\/$ about the z-axis of the spin, 
upto a global phase factor.  These z-rotations have
been implemented using composite-z pulses, whereby a rotation
by an arbitrary angle $\theta\/$ about the z-axis, can be
decomposed as a set of rotations about the x and y 
axes~\cite{ernst-book}:
\begin{equation}
(\theta)_{z} \equiv (\pi/2)_{x} (\theta)_{y} (\pi/2)_{-x}
\label{composite}
\end{equation}
Global phase changes are not detectable 
in NMR and are hence ignored.
The spectrum reflects the constant or balanced nature of
the function implemented (Fig.~\ref{onebit}). 
After the implementation of a balanced
function, the qubit is in a state 
out-of-phase with the rest of the spectrum. 
The modified D-J algorithm for one qubit demonstrates the
power inherent in even a single bit of quantum information.

\begin{figure}[h]
\unitlength=1mm 
\begin{picture}(80,60)
\put(0,3){\psfig{figure=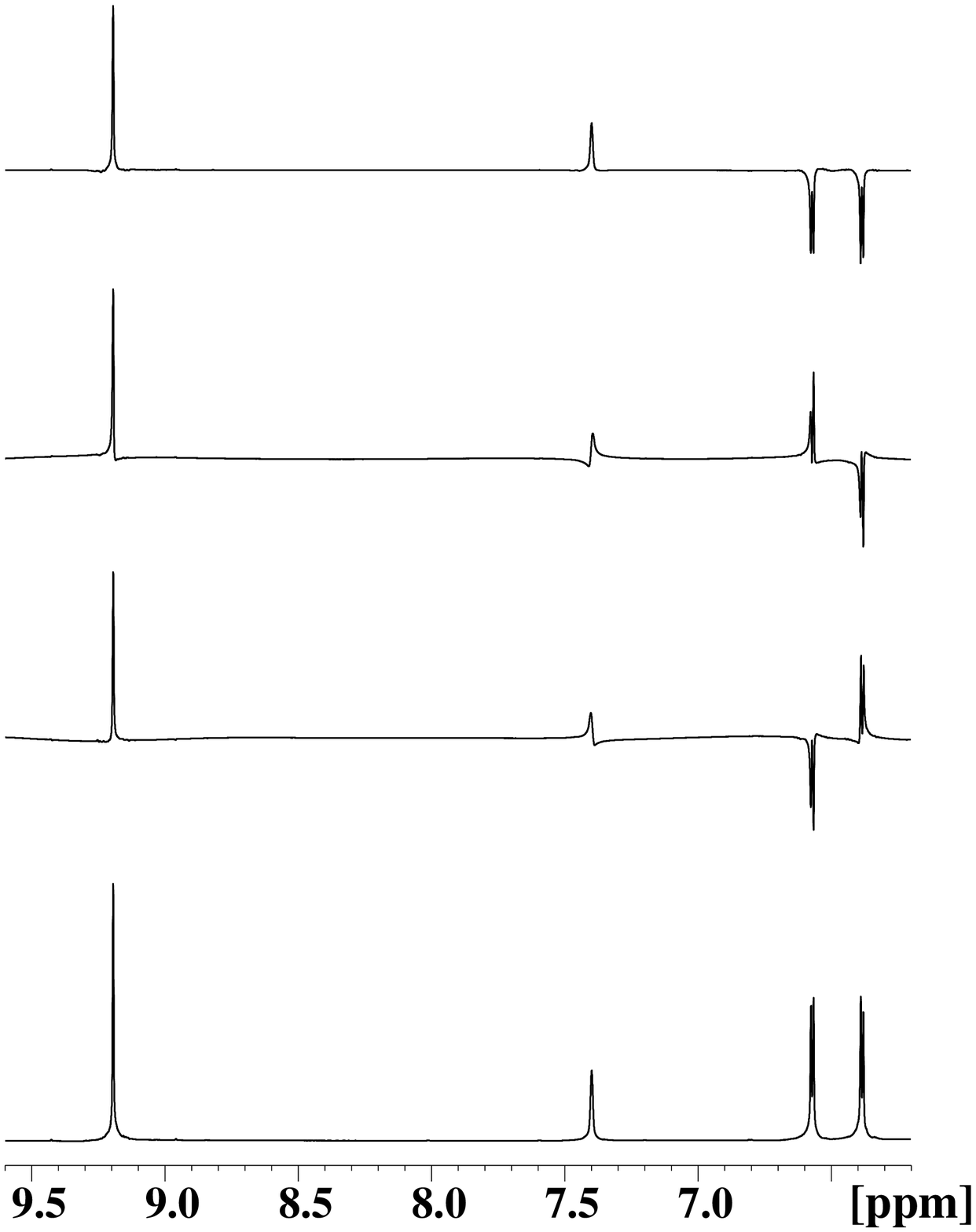,height=6cm,width=7cm,angle=0}}
\boldmath
\put(46,10){Constant} 
\put(70,10){$U^{\mbox{\tiny (2-bit)}}_{1}\/$}
\put(46,29){Balanced}
 \put(70,29){$U^{\mbox{\tiny (2-bit)}}_{2}\/$}
\put(46,43){Balanced} 
\put(70,43){$U^{\mbox{\tiny (2-bit)}}_{3}\/$}
\put(46,57){Balanced} 
\put(70,57){$U^{\mbox{\tiny (2-bit)}}_{4}\/$}
\unboldmath
\end{picture}
\hspace*{0.5cm}
\caption{The modified D-J algorithm for two qubits,
 implemented on 5-nitro-2-furaldehyde with the two qubits 
resonating at 6.47 ppm and 6.29 ppm respectively.}
\label{twobit}
\end{figure}

All the 8 unitary transformations corresponding to the 
functions for the two-qubit case are given in terms of
$I^{(j)}\/$ and $\sigma_z^{(j)}$ as
\begin{equation}
\begin{array}{ccll}
U^{\mbox{\tiny (2-bit)}}_{1} &= &I^{(1)} \otimes I^{(2)} &
\mbox{(Const.)}  \\
U^{\mbox{\tiny (2-bit)}}_{2} &= & 
\sigma^{(1)}_{z} \otimes I^{(2)} &
\mbox{(Bal.)}  \\
U^{\mbox{\tiny (2-bit)}}_{3} &=  &I^{(1)}
 \otimes  \sigma^{(2)}_{z} &
\mbox{(Bal.)}  \\
U^{\mbox{\tiny (2-bit)}}_{4} &= & \sigma^{(1)}_{z} 
\otimes  \sigma^{(2)}_{z}& 
\mbox{(Bal.)}  \\
U^{\mbox{\tiny (2-bit)}}_{i} &=& - U_{i-4} \,\, 
\mbox{(i=5,6,7,8)}& 
\end{array}
\end{equation}
All these operators are direct products of single spin operators.
They are thus incapable of generating entangled states and can
be implemented by operations on individual spins.
A pseudo-Hadamard transformation was performed 
on all the spins
(initially in thermal equilibrium) prior to the
execution of the desired $U_{i}\/$ transformations. 
The two constant functions correspond to the
``do-nothing'' operation, experimentally.
The NMR implementation of the balanced 
functions $U^{\mbox{\tiny (2-bit)}}_{2}\/$ and
$U^{\mbox{\tiny (2-bit)}}_{3}\/$  involves
rotations by $\pi\/$ about the z-axis in the single-spin
subspaces of spins 1 and 2 respectively, and have
been achieved using composite-z pulses (Eqn.~\ref{composite}).
The $U^{\mbox{\tiny (2-bit)}}_{4}\/$ transformation too, does 
not require the scalar J spin interaction and is 
implemented as successive $\pi\/$ rotations about 
the z-axes of spins 1 and 2 respectively.
The balanced functions are distinguished by one(or both) the
spins being out-of-phase with the rest of the NMR 
spectrum (Fig.~\ref{twobit}).
Only half the total number of functions have been
shown in the one and two-qubit cases, as the
others are merely negatives of these, and lead to
the same spectral patterns (spectra not shown).
The three-qubit D-J algorithm affords the simplest
example where quantum entanglement plays a
definitive role in the computation. The task here
is to implement $72\/$ unitary transformations, the
explicit operator forms for $9\/$ of which are 
\begin{eqnarray}
&&
\begin{array}{ccll}
U^{\mbox{\tiny (3-bit)}}_{1} &=& I^{(1)} 
\otimes I^{(2)} \otimes I^{(3)} 
&\mbox{(Const.)} \\
U^{\mbox{\tiny (3-bit)}}_{2} &=& 
\sigma^{(1)}_{z} \otimes I^{(2)} \otimes I^{(3)} 
&\mbox{(Bal.)}  \\
U^{\mbox{\tiny (3-bit)}}_{3} &=& 
I^{(1)} \otimes I^{(2)} \otimes  \sigma^{(3)}_{z}
&\mbox{(Bal.)} \\
U^{\mbox{\tiny (3-bit)}}_{4} &=& 
\sigma^{(1)}_{z} \otimes  \sigma^{(2)}_{z} \otimes I^{(3)}
&\mbox{(Bal.)} \\
U^{\mbox{\tiny (3-bit)}}_{5} &=& 
\sigma^{(1)}_{z} \otimes  \sigma^{(2)}_{z} 
\otimes  \sigma^{(3)}_{z}
&\mbox{(Bal.)} 
\end{array} \nonumber\\
&&\begin{array}{ccrl}
U^{\mbox{\tiny (3-bit)}}_{6} &=& \frac{1}{2}
\sigma^{(1)}_{z}
 \otimes ( I^{(2)} \otimes I^{(3)} +
\sigma^{(2)}_{z} \otimes I^{(3)} + &\\ 
&& I^{(2)} \otimes \sigma^{(3)}_{z}
- \sigma^{(2)}_{z} \otimes \sigma^{(3)}_{z} ) 
& \mbox{(Bal.)} \\
U^{\mbox{\tiny (3-bit)}}_{7} &=& 
\frac{1}{2}
\sigma^{(2)}_{z} \otimes ( I^{(1)} \otimes I^{(3)} +
\sigma^{(1)}_{z} \otimes I^{(3)} +&  \\ 
&& I^{(1)} \otimes \sigma^{(3)}_{z}
- \sigma^{(1)}_{z} \otimes \sigma^{(3)}_{z} ) 
& \mbox{(Bal.)} \\
U^{\mbox{\tiny (3-bit)}}_{8} &=& \frac{1}{2}
\sigma^{(3)}_{z} 
\otimes ( I^{(1)} \otimes I^{(2)} +
\sigma^{(1)}_{z} \otimes I^{(2)} + & \\ 
&& I^{(1)} \otimes \sigma^{(2)}_{z}
- \sigma^{(1)}_{z} \otimes \sigma^{(2)}_{z})
& \mbox{(Bal.)}  
\end{array} \nonumber\\
&&U^{\mbox{\tiny (3-bit)}}_{9} =\frac{1}{2}(
\sigma^{(1)}_{z} \otimes I^{(2)} \otimes I^{(3)}
+ I^{(1)} \otimes I^{(2)} \otimes \sigma^{(3)}_{z} -
\nonumber \\
&&\quad\quad\,\,  \sigma^{(1)}_{z}
 \otimes \sigma^{(2)}_{z} \otimes I^{(3)}
+  I^{(1)} \otimes \sigma^{(2)}_{z} 
\otimes \sigma^{(3)}_{z})\,\, 
\mbox{(Bal.)} \!\!\!
\end{eqnarray}
The operators $U^{\mbox{\tiny (3-bit)}}_{1}\/$ - 
$U^{\mbox{\tiny (3-bit)}}_{5}\/$ can be decomposed as
direct products of single-spin operators and
are thus non-entangling transformations. 
The operators $U^{\mbox{\tiny (3-bit)}}_{6}\/$ 
- $U^{\mbox{\tiny (3-bit)}}_{9}\/$ cannot be
decomposed as direct products of single-spin
operators and are hence capable of generating
entangled states from non-entangled ones. 
The operators $U^{\mbox{\tiny (3-bit)}}_{6}\/$, 
$U^{\mbox{\tiny (3-bit)}}_{7}\/$, and
$U^{\mbox{\tiny (3-bit)}}_{8}\/$, are entangling in 
different two-spin subspaces  and can be
factored as direct products of a single-spin
operator and a two-particle entangling
transformation.  They can thus generate states in
which two qubits are entangled, with the
third qubit remaining non-entangled with
either of them.   On the other hand, the
transformation $U^{\mbox{\tiny (3-bit)}}_{9}\/$ 
does not allow any such simplifications. It
is maximally entangling and leads to states that are
three-qubit entangled.   
These $9\/$ functions are thus divided into
three categories namely, non-entangling,
two-qubit entangling, and maximally (three-qubit)
entangling. The remaining functions are similar
in form, and can be classed into one or 
the other of these categories. 

\begin{figure}[n]
\unitlength=1mm 
\begin{picture}(80,80)
\put(5,0){\psfig{figure=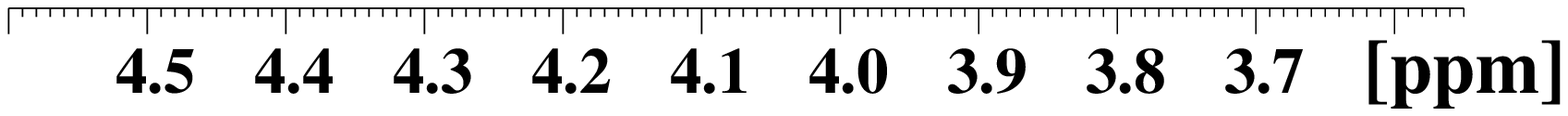,height=2.5cm,width=7cm,angle=0}}
\put(0,0){\psfig{figure=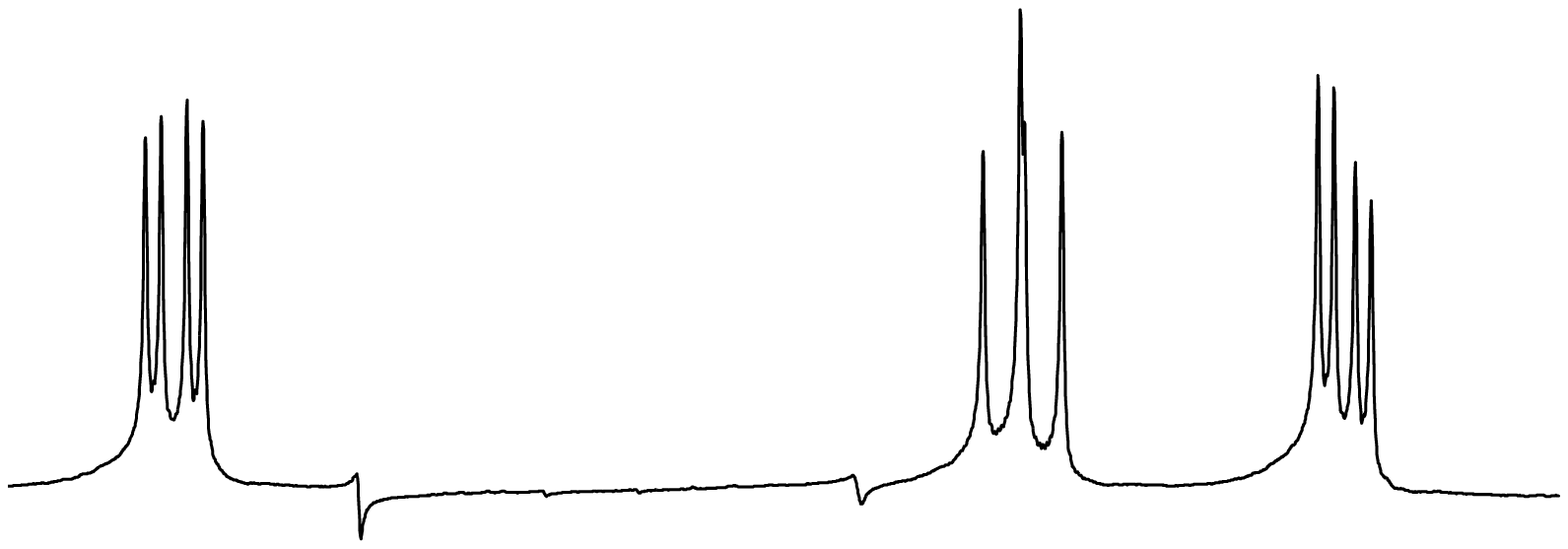,height=1.5cm,width=7cm,angle=0}}
\boldmath
\put(20,7){Constant} \put(65,7){$U^{\mbox{\tiny (3-bit)}}_{1}\/$}
\put(0,15){\psfig{figure=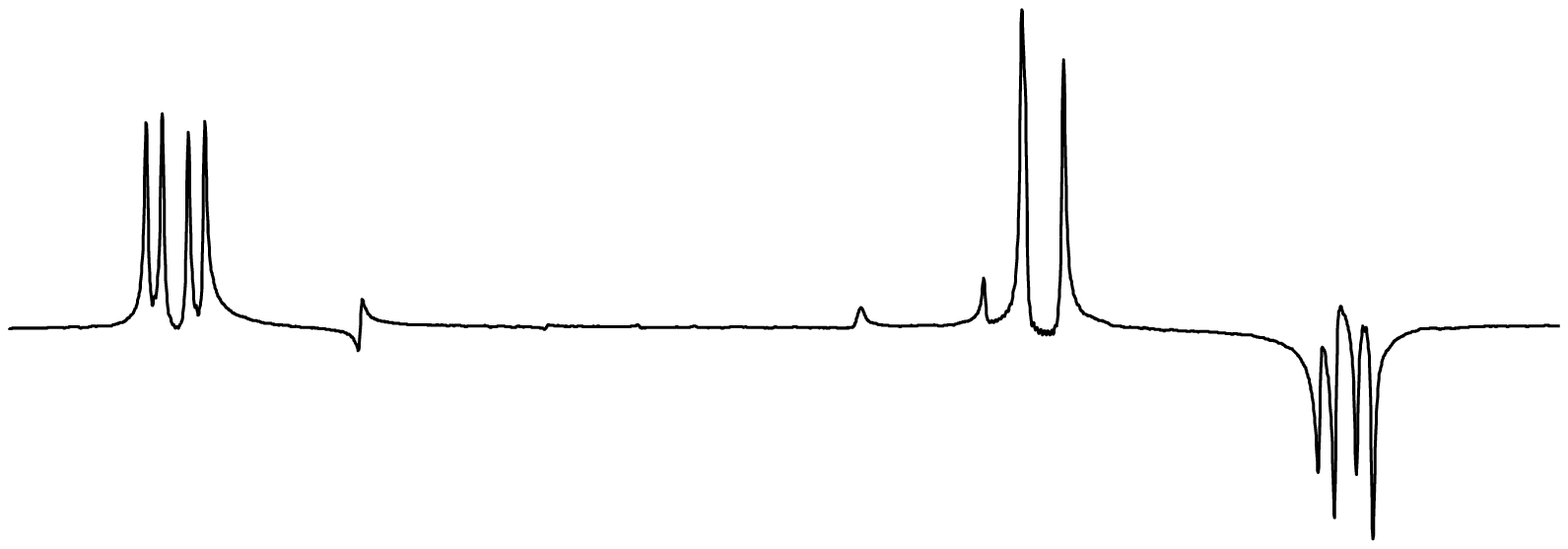,height=1.5cm,width=7cm,angle=0}}
\put(20,25){Balanced} \put(65,25){$U^{\mbox{\tiny (3-bit)}}_{2}\/$}
\put(0,30){\psfig{figure=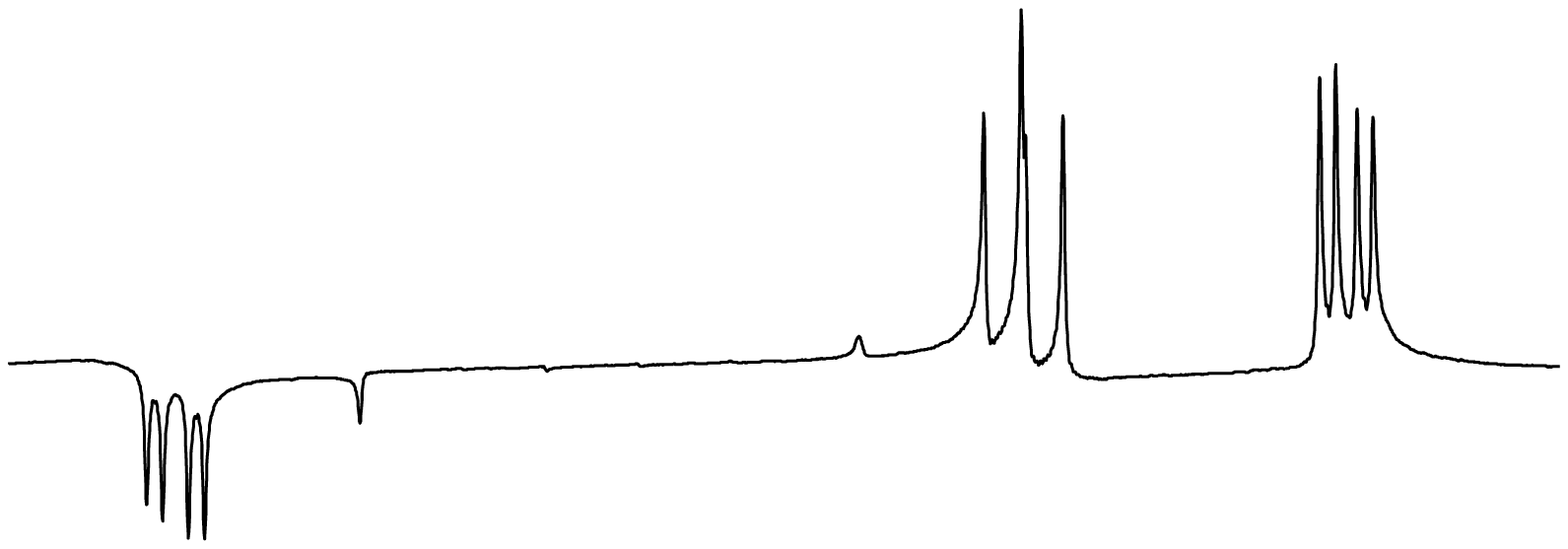,height=1.5cm,width=7cm,angle=0}}
\put(20,40){Balanced} \put(65,40){$U^{\mbox{\tiny (3-bit)}}_{3}\/$}
\put(0,45){\psfig{figure=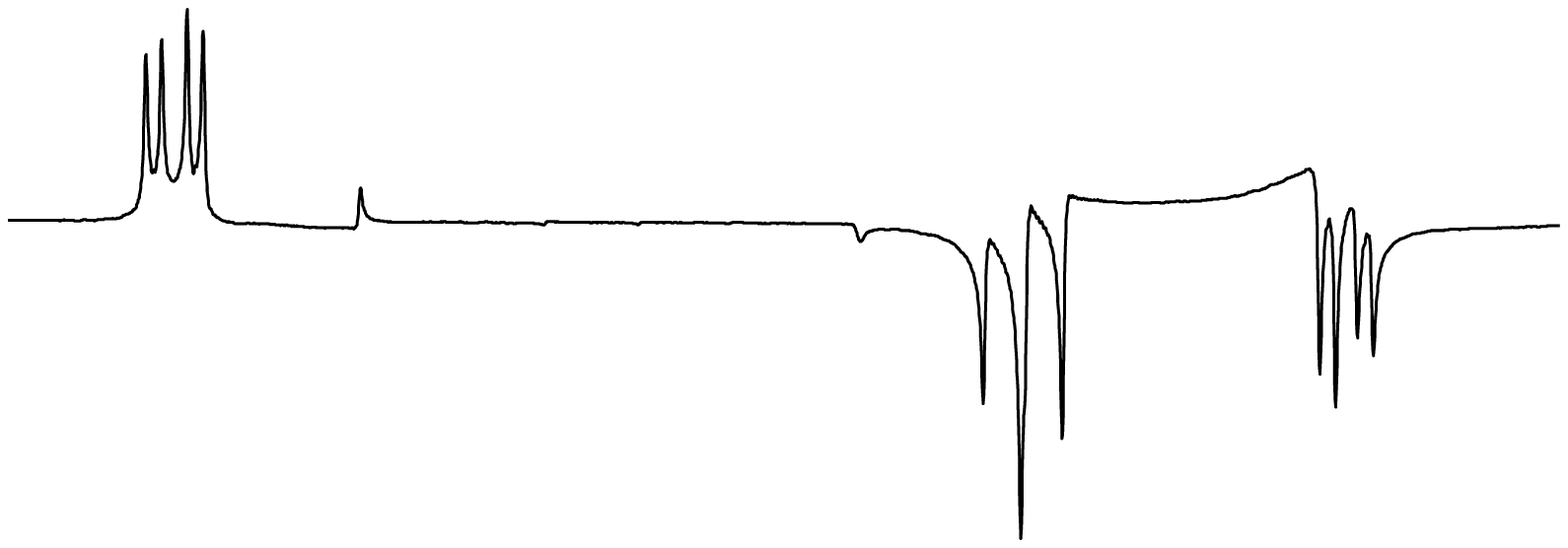,height=1.5cm,width=7cm,angle=0}}
\put(20,56){Balanced} \put(65,56){$U^{\mbox{\tiny (3-bit)}}_{4}\/$}
\put(0,60){\psfig{figure=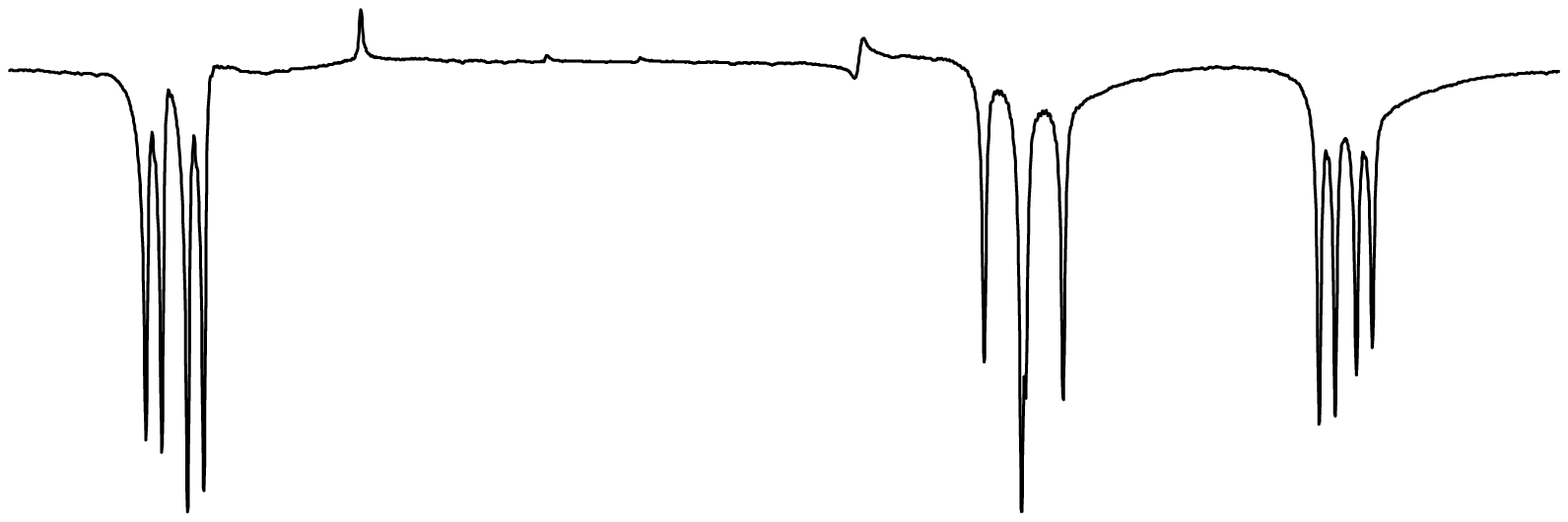,height=1.5cm,width=7cm,angle=0}}
\put(20,76){Balanced} \put(65,76){$U^{\mbox{\tiny (3-bit)}}_{5}\/$}
\unboldmath
\end{picture}
\vspace*{1cm}
\caption{The refined D-J algorithm for three qubits, implemented
on 2,3-dibromopropionic acid. The functions shown are all 
non-entangling in nature.}
\label{threebit1}
\end{figure}

The result of experimentally
applying the non-entangling 
transformations  $U^{\mbox{\tiny (3-bit)}}_{1}\/$ -
$U^{\mbox{\tiny (3-bit)}}_{5}\/$, 
after a pseudo-Hadamard transformation
on a thermal equilibrium state, 
is shown in Figure~\ref{threebit1}. The constant
function $U^{\mbox{\tiny (3-bit)}}_{1}\/$ 
is the unit operator,  and corresponds to the
``no pulse'' or the ``do-nothing'' operation. The
balanced functions $U^{\mbox{\tiny (3-bit)}}_{2}\/$
and $U^{\mbox{\tiny (3-bit)}}_{3}\/$ correspond
to a rotation by the angle $\pi\/$ about the
$z$-axis of the first and the third spins
respectively, without perturbing the other spins.
This has been achieved by a spin-selective
$[\pi]_z\/$ pulse in each case, using
composite $z$-pulses (Eqn.~\ref{composite}). The
spectrum is categorised by the spin in question
being out-of-phase with the rest of the spectrum.
The transformation $U^{\mbox{\tiny (3-bit)}}_{4}\/$ has
been implemented by two spin-selective 
$[\pi]_{z}\/$ pulses applied consecutively on
the first and the second spins respectively, and leads
to a spectrum with both these spins being out-of-phase
with the third. The non-entangling balanced
function $U^{\mbox{\tiny (3-bit)}}_{5}\/$, has been
implemented by successive spin-selective 
$[\pi]_{z}\/$ pulses on all the three spins. 

\begin{figure}
\unitlength=1mm 
\begin{picture}(80,60)
\boldmath
\put(5,0){\psfig{figure=ppm.eps,height=2.5cm,width=7cm,angle=0}}
\put(0,0){\psfig{figure=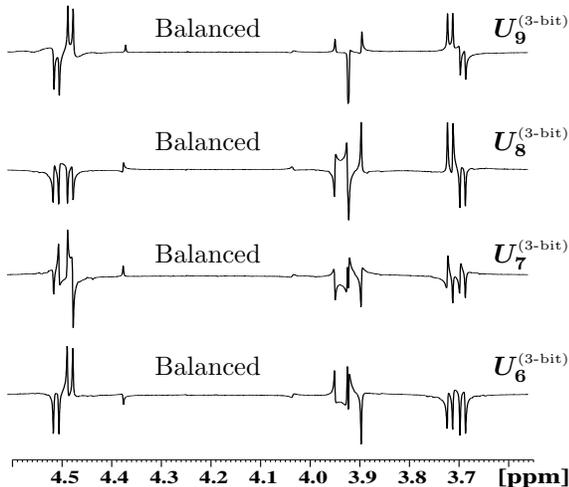,height=6cm,width=7cm,angle=0}}
\put(20,10){Balanced} \put(65,10){$U^{\mbox{\tiny (3-bit)}}_{6}\/$}
\put(20,25){Balanced} \put(65,25){$U^{\mbox{\tiny (3-bit)}}_{7}\/$}
\put(20,40){Balanced} \put(65,40){$U^{\mbox{\tiny (3-bit)}}_{8}\/$}
\put(20,55){Balanced} \put(65,55){$U^{\mbox{\tiny (3-bit)}}_{9}\/$}
\unboldmath
\end{picture}
\vspace*{0.5cm}
\caption{Entangling balanced functions implemented  on the
three-qubit system of 2,3-dibromopropionic acid.}
\label{threebit3}
\end{figure}

The two-qubit entangling transformation 
$U^{\mbox{\tiny (3-bit)}}_{6}\/$, 
is achieved by the pulse sequence
$$
\left[\pi/2\right]^{2}_{z} 
\,\,
\left[\pi/2\right]^{3}_{z}
\stackrel{(\tau_{23}/2)}{-\!\!\!-\!\!\!-\!\!\!-}
\left[\pi\right]^{2}_{x}
\,\,
\left[\pi\right]^{3}_{x}
\stackrel{(\tau_{23}/2)}{-\!\!\!-\!\!\!-\!\!\!-}
\left[\pi\right]^{1}_{z}
$$
where $\tau_{23} = 1/J_{23}\/$ and 1,2 and 3 are qubit labels. 
The operators $U^{\mbox{\tiny (3-bit)}}_{7}\/$ 
and $U^{\mbox{\tiny (3-bit)}}_{8}\/$  
correspond to cyclic permutations of 
the qubits.   The spin-selective
$\pi\/$ pulses in the middle of the
free evolution period $\tau_{23}\/$ refocus the
chemical shift evolution. The pulse sequence 
(applied after a pseudo-Hadamard transformation 
on all three qubits in a thermal initial state) 
results in a density matrix with the product
operator form
$ - I^{1}_{x} + 2 I^{2}_{x} I^{3}_{z} +
2 I^{2}_{z} I^{3}_{x}\/$, 
leading to a spectrum with the multiplet of the
first qubit inverted, and an antiphase
doublet of doublet pattern for 
the other two qubits (Fig.~\ref{threebit3}). 
The three-qubit entangling function 
$U^{\mbox{\tiny (3-bit)}}_{9}\/$ is
implemented by the pulse sequence 
$$
\left[\pi/2\right]^{1}_{z}
\left[\pi/2\right]^{3}_{z}
\stackrel{(\tau_{12}/2)}{-\!\!\!-\!\!\!-\!\!\!-}
\left[\pi\right]^{1}_{x}
\left[\pi\right]^{2}_{x}
\stackrel{(\tau_{12}+\tau_{23})/2}{
-\!\!\!-\!\!\!-\!\!\!-\!\!\!-\!\!\!-}
\left[\pi\right]^{2}_{x}
\left[\pi\right]^{3}_{x}
\stackrel{(\tau_{23}/2)}{-\!\!\!-\!\!\!-\!\!\!-}
$$
leading to an antiphase spectral pattern
for all three qubits, corresponding to
the product operators
$ 2 I^{1}_{x} I^{2}_{z} + 2 I^{2}_{z} I^{3}_{x} +
4 I^{1}_{z} I^{2}_{x} I^{3}_{z}\/$ (Fig.~\ref{threebit3}). 
The spectra in Figure~\ref{threebit3} 
suffer from  phase distortions arising from the
inaccurate refocusing of the chemical shifts during
the $\tau\/$ periods, and J-evolution during long
spin-selective composite-z pulses 
($[\pi]^{i}_{z} \equiv [\pi/2]^{i}_{x} [\pi]^{i}_{y} 
[\pi/2]^{i}_{-x} = 42\/$ msec).
This implementation of the D-J algorithm 
does not require the initial preparation of
the spins in a pseudo-pure state, since the thermal
equilibrium state serves equally well as a good initial state.
The observable spectral result is the same in both cases, though
beginning with a pseudo-pure state creates some 
(undetectable) multiple-quantum coherences.
The final pseudo-Hadamard transformation to
extract the constant or balanced nature of
the function (Fig.~\ref{working}) is 
canceled by the $(90)^{0}\/$ read-out
pulse usually used in NMR experiments, and the
computation essentially culminates in the
application of the desired $U_f\/$ function after
the first pseudo-Hadamard transformation.
  
A modification to the usual D-J algorithm 
enabled an $n\/$-bit implementation using
$n\/$ qubits. The required unitary transformations
were tailored to eliminate the need for the
extra qubit, and the modified D-J algorithm was
tested experimentally for one, two and three qubits.
While the one and two qubit cases use non-entangling
unitary transformations, it was noted that 
for three (or more) qubits, 
multi-particle entangling transformations 
are required.\\
\leftline{\bf Acknowledgement:} The use of the
AMX-400 spectrometer at SIF, IISc Bangalore, funded
by DST New Delhi, is gratefully acknowledged. 

\end{document}